\documentclass[epj]{webofc}
\usepackage[varg]{txfonts}   % Web of Conferences font

\usepackage{amsmath}
\usepackage{amsfonts}
\usepackage{amssymb}
\newcommand{\sqrts}{\sqrt{s}}

\newcommand{\sqrtsgzk}{\sqrt{s_{_{\rm GZK}}}}
\newcommand{\pp}{p-p}
\newcommand{\epos}{\textsc{epos}}
\newcommand{\qgsjet}{\textsc{qgsjet}} 
\newcommand{\sibyll}{\textsc{sibyll}}
\newcommand{\dNdeta}{\rm dN_{ch}/d\eta|_{\eta=0}}

\newcommand{\Xmax}{X$_{\rm max}$}
\newcommand{\ECR}{E$_{_{\rm CR}}$}

\woctitle{International Conference on New Frontiers in Physics 2013}

\begin{document}
%
%\title{Particle physics at the collider and cosmic frontiers: Status after LHC Run 1}
%\title{Impact of LHC Run-1 results on particle astrophysics}
\title{Impact of LHC Run-1 on particle astrophysics}
%\title{Collider and cosmic frontiers: Connections after LHC Run 1}
%%%\subtitle{???}

\author{David d'Enterria\inst{1}\fnsep\thanks{\email{dde@cern.ch}}}

\institute{CERN, PH Department, CH-1211 Geneva 23, Switzerland}

\abstract{An overview of the impact of the first three years of LHC operation on two of the most important open
  questions in astroparticle physics is presented. Measurements in proton-proton collisions at the energy
  frontier that provide valuable information on the identity of the highest-energy particles in the cosmos
  as well as new constraints on the nature of dark matter, are summarized.} %pervading the universe.}
\maketitle
%

%%%%%%%%%%%%%%%%%%%%%%%%%%%%%%%%%%%%%%%%%%%%%%%%%%%%%%%%%%%%%%%%%%%%%%%%%%%%%%%%%%%%%%%%%%%%%%%%%%%%%%%%%
%
\section{Introduction}
\label{intro}

The nature of the dark matter that pervades the universe~\cite{Bauer:2013ihz} as well as the identity and origin
of the highest-energy cosmic rays reaching Earth~\cite{Engel:2011zzb} are among the most pressing open
questions in particle astrophysics today. During the last 3 years, the CERN Large Hadron Collider (LHC) has
been colliding protons (and nuclei) at the highest center-of-mass energies ever studied in the laboratory
(up to $\sqrts$~=~8~TeV). Many measurements carried out by the different experiments provide
new insights on the solution of both open issues. On the one hand, detailed LHC studies of the theory of the
strong interaction (quantum chromodynamics, QCD) have resulted in valuable constraints on the Monte Carlo
models used to interpret the extended air-showers produced by ultra-high-energy cosmic rays (UHECR) colliding
with air nuclei in the upper atmosphere. On the other, searches of dark matter (DM) particles --which escape
the LHC detectors but can be discovered as an excess of events with missing transverse energy-- have been
carried out in different final-states corresponding to various visible particles accompanying DM
production. The latest developments in UHECR and DM physics as well as the most relevant LHC measurements
providing information on both key problems are summarized hereafter.

%%%%%%%%%%%%%%%%%%%%%%%%%%%%%%%%%%%%%%%%%%%%%%%%%%%%%%%%%%%%%%%%%%%%%%%%%%%%%%%%%%%%%%%%%%%%%%%%%%%%%%%%%
%
\section{Impact of LHC data on ultra-high-energy cosmic rays}
\label{sec:UHECR_LHC}

\begin{figure}
%\centering
\includegraphics[width=0.62\textwidth,clip]{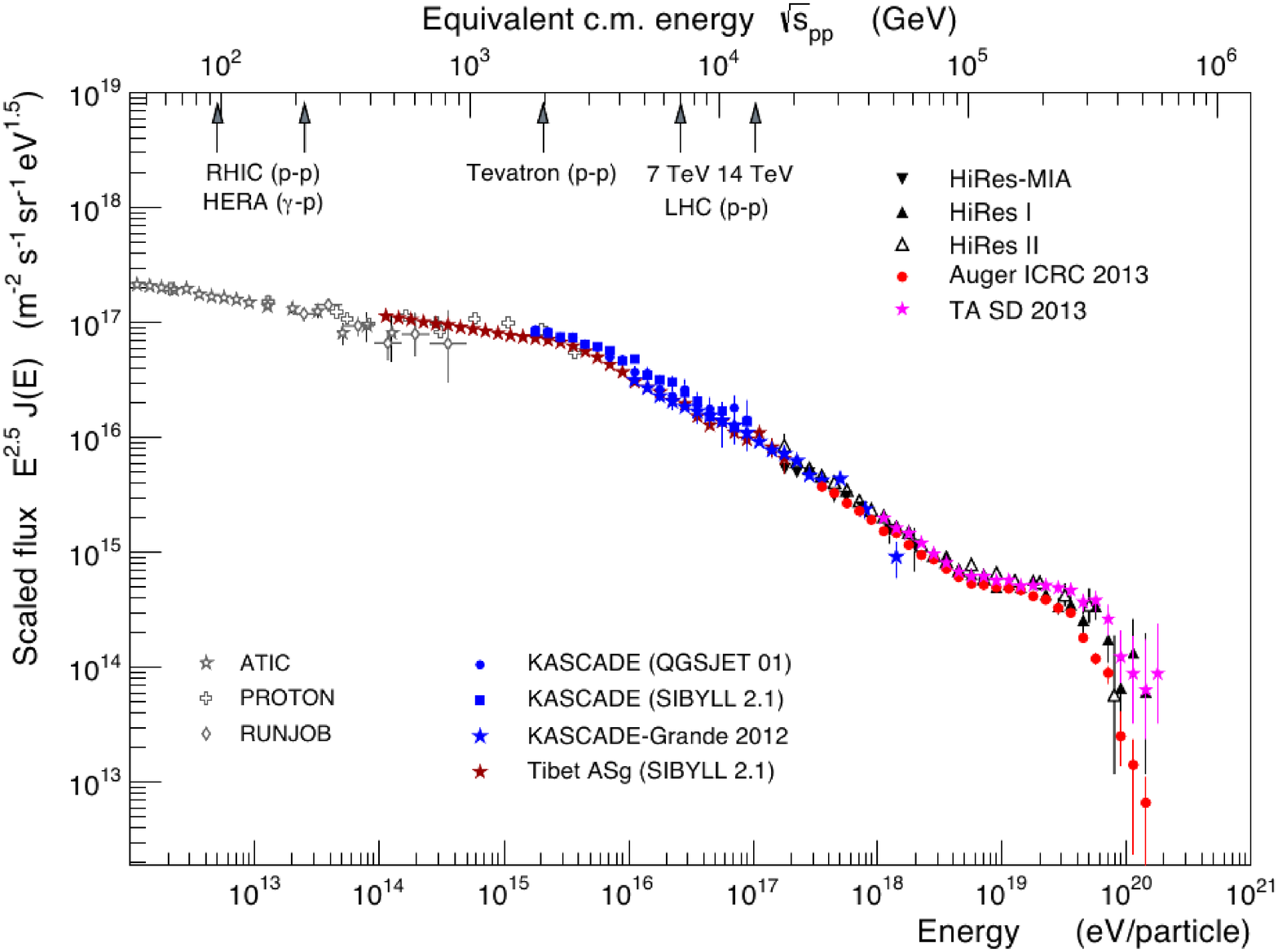}\hspace{0.2cm}
\includegraphics[width=0.37\textwidth,height=6.cm,clip]{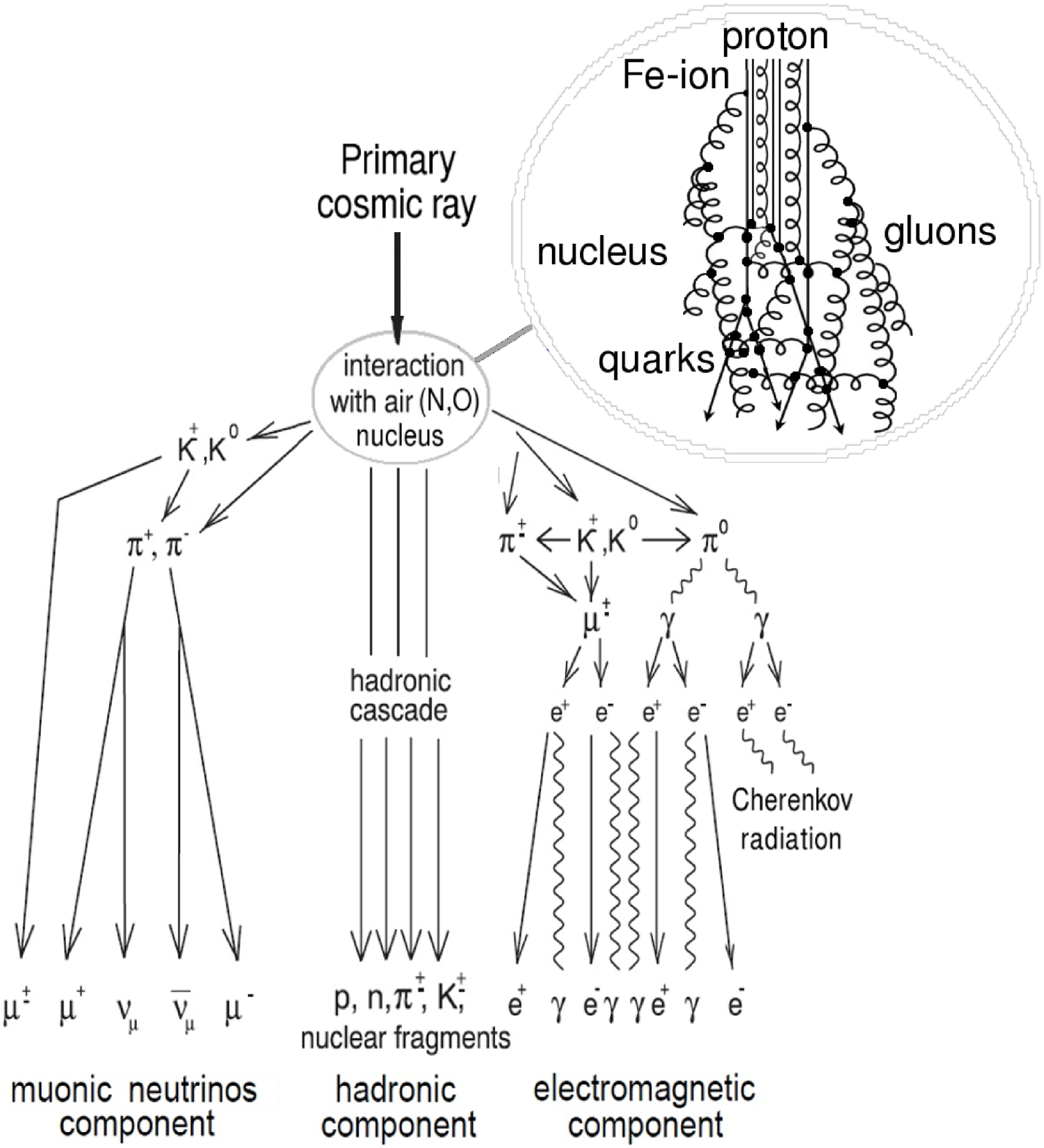}
\caption{Left: Cosmic rays flux (scaled by \ECR$^{\!2.5}$) as a function of CR energy (equivalent %collider
  c.m.~energies for various colliders are shown on the top axis)~\cite{Pierog:2013dya}. 
  Right: Schematic ``microscopic'' view of an extensive air shower produced by an ultrarelativistic cosmic ray 
  (proton or Fe-ion) colliding with a nucleus in the upper atmosphere.} 
\label{fig:CR1}
\end{figure}

The highest energy hadronic interactions on Earth occur in collisions of cosmic rays
--protons and nuclei produced in various astrophysical sources-- with air nuclei as they enter the
atmosphere~\cite{Engel:2011zzb}. The highest cosmic-ray energies measured are of the order of
$10^{20}$~eV (Fig.~\ref{fig:CR1}, left), corresponding to the Greisen-Zatsepin-Kuzmin (GZK) cutoff 
of CRs propagating through intergalactic space and dissociating in collisions with the microwave
background~\cite{GZK} and/or to the maximum energy reachable in astrophysical accelerators~\cite{Hillas:1985is}.
The experimental determination of the primary UHECR energy and mass relies upon the comparison of the 
properties of the produced extensive air-showers (EAS, Fig.~\ref{fig:CR1} right) %that they produce %in the atmosphere  
 --such as the shower peak position \Xmax, and the number of electrons and muons on ground N$_{\rm e,\mu}$--
to Monte Carlo (MC) hadronic simulations: \epos~\cite{Werner:2005jf,Pierog:2013ria},
\qgsjet01~\cite{Kalmykov:1997te}, \qgsjet-II~\cite{Ostapchenko:2010vb} and \sibyll~\cite{Ahn:2009wx}. 
The dominant source of uncertainty in the interpretation of the EAS data stems from our
limitations to model particle production in strongly-interacting systems at c.m.~energies up to
$\sqrtsgzk\approx 400$\,TeV, i.e. more than two orders of magnitude higher than those studied at particle
colliders before the LHC. Indeed, even at asymptotically high energies the collision between two hadronic
objects is sensitive to non-perturbative (hadronization, beam remnants, soft peripheral diffractive
scatterings) as well as semihard (saturation of gluon densities, multiparton interactions)
dynamics that need to be directly constrained  from experimental data~\cite{d'Enterria:2011kw}.\\

\begin{figure}
%\centering
\includegraphics[width=0.48\textwidth,height=5.cm,clip]{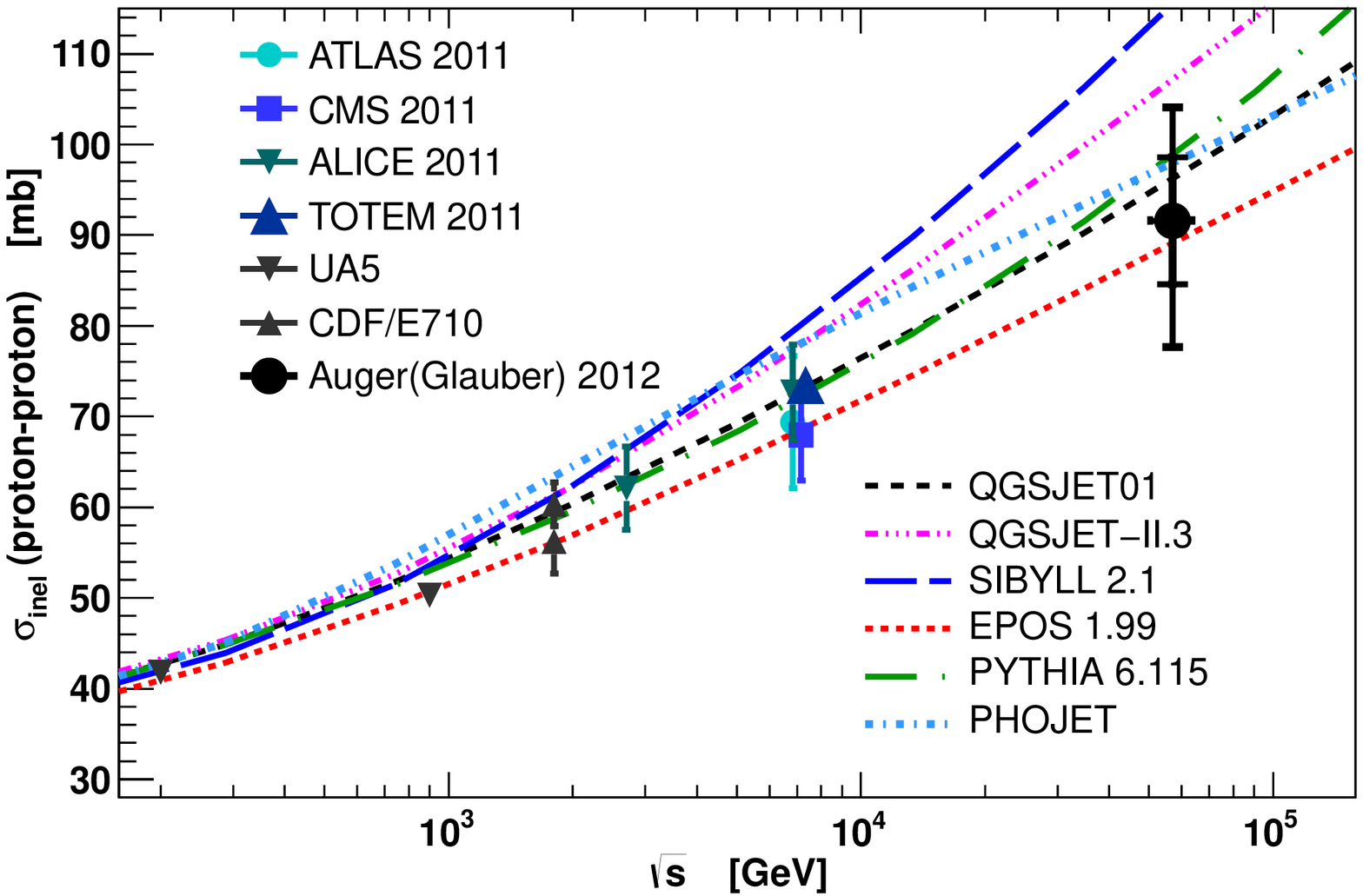}
\includegraphics[width=0.52\textwidth,height=5.cm,clip]{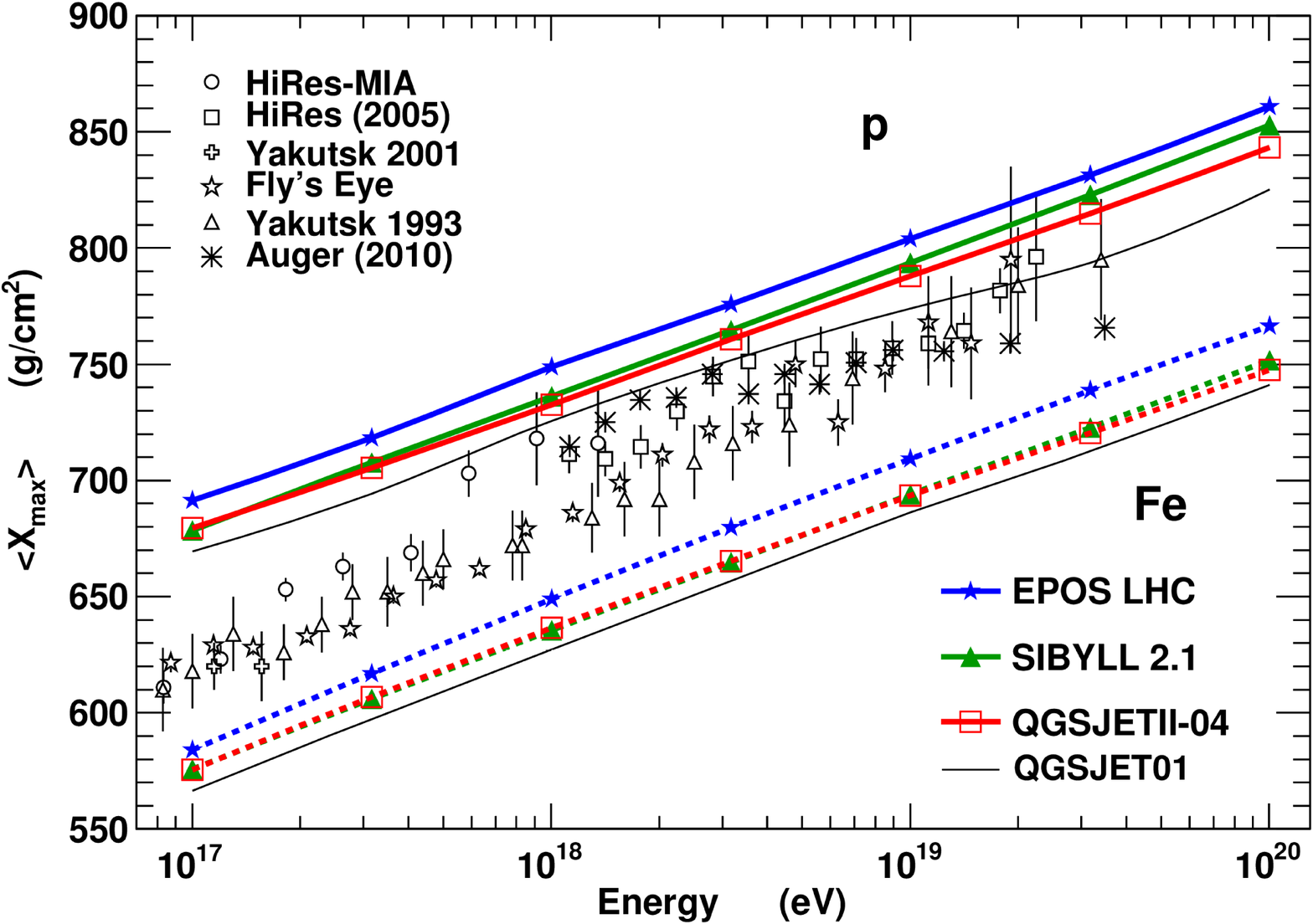}
\caption{Left: Inelastic \pp\ cross sections measured at colliders and estimated by
  the P.~Auger Observatory at $\sqrts\approx$~50~TeV~\cite{Auger:2012wt}, compared to hadronic MC
  predictions. Right: Measured mean \Xmax\ position as a function of CR energy, compared to hadronic MC
  predictions for proton- and Fe-induced showers in the atmosphere~\cite{Pierog:2013dya}.}  
\label{fig:CR2}
\end{figure}

The LHC has extended by more than a factor of three the c.m.~energies for which we have 
direct p-p measurements available, going beyond the ``knee'' structure of the CR spectrum at 
\ECR$\approx$~10$^{15.5}$~eV (Fig.~\ref{fig:CR1}, left). The following LHC inclusive observables %measured at the LHC are
are sensitive to the non-perturbative and semihard QCD dynamics %and help constrain the ingredients of the 
implemented in hadronic MCs commonly used in UHECR physics:
\begin{itemize}
%\begin{description}
\item Inelastic \pp\ cross section $\sigma_{\rm inel}$. Hadronic cross sections are not directly
computable from the QCD Lagrangian\footnote{Although the possibility of using lattice QCD
calculations~\cite{Giordano:2012mn} one day, should not be discarded.}, 
but are constrained by basic quantum mechanical relations (such as 
the Froisart bound, the optical theorem, and dispersion relations) which can be combined with experimental
data to make predictions. The measured inelastic \pp\ cross section, 
$\sigma_{\rm inel(visible)}$~=~73~(60)~$\pm$~2~mb at
$\sqrts$~=~7~TeV~\cite{Antchev:2013iaa,Aad:2011eu,Chatrchyan:2012nj,Abelev:2012sea}, 
was mostly overpredicted by the MCs (Fig.~\ref{fig:CR2}, left), 
which tended to over (under) estimate the diffractive contributions at high (low) masses.
%shows the LHC measurements (together with the value estimated at $\sqrts\approx$~50~TeV by Auger~\cite{Auger:2012wt}) compared to the
%different theoretical curves. 
The measured value of $\sigma_{\rm inel}$ at the LHC implies a reduced $\sigma_{\rm inel}$(p-Air) cross
section and subsequently a deeper \Xmax\ shower position of UHECR.

\item Pseudorapidity density of charged particles at midrapidity $\dNdeta$ and event-by-event
distribution of the charged particle multiplicity $\rm P(N_{\rm ch})$. At LHC energies about 70\% of the produced
hadrons come from (multi)parton interactions (with exchanged momenta of order 1--2~GeV) followed by
their fragmentation into hadrons. % Although the 900-GeV data well reproduced (MCs tuned to SppS, Tevatron)
The measurements of $\dNdeta$ and $\rm P(N_{\rm ch})$ have been crucial to improve the modeling of multiparton interactions
(at large values of $\rm N_{\rm ch}$) and diffraction (for low $\rm N_{\rm ch}$). None of the pre-LHC CR hadronic models predicted
precisely both distributions, although taken together they ``bracketed'' the experimental data~\cite{d'Enterria:2011kw}.

\item  Energy distribution of (very) forward particles. 
%$dE_{had}/d\eta|_{|\eta|=3-5}$~\cite{Chatrchyan:2011wm,Aaij:2012pda,Aspell:2012ux} and 
%$dN/dE_{\gamma}|_{|\eta|>10.94}$~\cite{Adriani:2011nf}. 
Knowledge of the particle and energy flows emitted at very small polar angles in hadronic collisions 
--dominated by underlying event activity, beam remnants and diffractive fragments-- is also 
crucial for understanding the first stages of the EAS development.  
Most of the UHECR models reproduced appropriately the data at pseudorapidities 
$|\eta|$~=~3--5~\cite{Chatrchyan:2011wm,Aaij:2012pda,Aspell:2012ux}, but less so
for the neutral activity close to the beam rapidity~\cite{Adriani:2011nf}.
%A significant fraction of the particles produced in hadronic interactions issue from the
%fragmentation of ``beam remnants'' at very small polar angles. hadron-hadron interactions 
%Important influence on cosmic-ray EAS development:
%Leading baryon (inelasticity) \& had-to-e.m. energy transfer (pi0 to gamma gamma)
%\item (iv) average transverse momentum of the produced hadrons $\meanpt$. %and
%<pT> is sensitive to pQCD x-sections & gluon-saturation
% <pT> should approach asymptotically the saturation scale: <pT> ~ Qsat
%CRs MCs predict very slow <pT> increase (but EPOS, due to collective flow)
% At GZK: <pT> ~ 0.6 ~ 1.0 GeV/c (PYTHIA: <pT> ~ 0.7 ~ 1.5 GeV/c)
%\end{description}
\end{itemize}

Two main conclusions can be extracted from the confrontation of data to pre-LHC
MCs~\cite{d'Enterria:2011kw}. First, although no model reproduced consistently all results, when taken
together they ``bracketed'' the LHC data and could be globally trusted to interpret the EAS properties. 
Second, the absence of a drastic change in the properties of multiparticle production at the LHC (corresponding to 
energies \ECR$\approx$~10$^{16}$~eV), confirmed that the CR ``knee'' at 10$^{15.5}$~eV is not due to the production of new
(unobserved) particles, as speculated in some cases, but to the change from a light to a more heavy mass CR composition.
The update of the event generators %\epos-LHC and \qgsjet-II-4, 
to account for the LHC measurements
%: Diffraction, MPI, saturation retuned. Extra MCs constraints coming from recent p-Pb @ 5 TeV (Feb'13).
has led to a convergence of their predictions: the \Xmax\ slope has decreased (increased) for \epos\ (\qgsjet-II) and 
%yielding an overall increase of the values of \Xmax\ and 
their N$_{\mu}$ values have raised at the highest CR energies (although they are still below the measured N$_{\mu}$).
%re-tuned model converge to old Sibyll 2.1 predictions
%-- %and a reduction of the overall uncertainties on the cosmic ray mass composition at the highest CR energies. 
% increase (converge with Sybill 2.1).
% Nmu increase (converge with EPOS)
%still ~30\% deficit compared to data
Around the GZK cutoff, the CR data prefer a mixed proton-iron composition (Fig.~\ref{fig:CR2}, right), with
reduced model uncertainties (from $\sim$50~g/cm$^2$ to $\sim$20~g/cm$^2$ for values of \Xmax\ where 
the proton-iron difference is about 100 g/cm$^2$)~\cite{Pierog:2013dya}.

%%%%%%%%%%%%%%%%%%%%%%%%%%%%%%%%%%%%%%%%%%%%%%%%%%%%%%%%%%%%%%%%%%%%%%%%%%%%%%%%%%%%%%%%%%%%%%%%%%%%%%%%%
%
\section{Impact of LHC data on dark matter searches}
\label{sec:DM_LHC}

Dark matter is required to explain many astrophysical and cosmological observations inferred from its
gravitational effects on visible matter, radiation, and the large-scale structure of the universe. 
The astronomical evidences for DM have grown steadily since
(i) the observed faster-than-Kleperian rotation curves of stars in the Milky Way (in the 1930s)
and, in general, in spiral galaxies (in the late 1970s); strongly supported by
(ii) the collision of galaxies in the Bullet Cluster~\cite{Markevitch:2003at}, where gravitational-lensing
reveals interaction-less galaxies lying ahead of the visible collisional gas; and more recently 
among others %(e.g. hot-gas in clusters, growth of clusters, ...) 
by e.g. (iii) the relative motion between the Milky Way and Andromeda, which indicates a combined virial
mass M$_{_{\rm virial}}=$~3.2$\cdot$10$^{12}\times$M$_{\odot}$, ten times larger than that 
estimated from the visible stars and gas~\cite{vanderMarel:2012xp}.
%\end{description}
The existence of DM is also supported by many cosmological observations such as
%\begin{description} \item 
(i) the large-scale structure of the universe given by galaxy distribution surveys confronted to statistical
simulations~\cite{Hawkins:2002sg}; 
(ii) the peak heights of the cosmic microwave background measured by WMAP and Planck~\cite{Komatsu:2010fb,Ade:2013zuv}
(lower than that expected in the absence of DM); plus
(iii) other measures of matter density (baryon acoustic oscillations, big bang nucleosyntesis, supernova
distances, etc.).
The post-Planck global fit of the energy budget of the universe yields a 26.8\% DM content, yielding 
a local halo density $\rho_{\rm \textsc{dm}}\approx$~0.3~GeV/cm$^3$ with average speed 
$\overline{v}_{\rm \textsc{dm}}$ of a few hundred km/s, and a flux of 10$^{5}$~cm$^{-2}$s$^{-1}$ on Earth.\\
%which, for a 100-GeV DM-particle $\chi$, corresponds to average speeds $\mean{v}\approx$~220~km/s.\\
%\end{description}

\begin{figure}
%\centering
\includegraphics[width=0.46\textwidth,height=5.cm,clip]{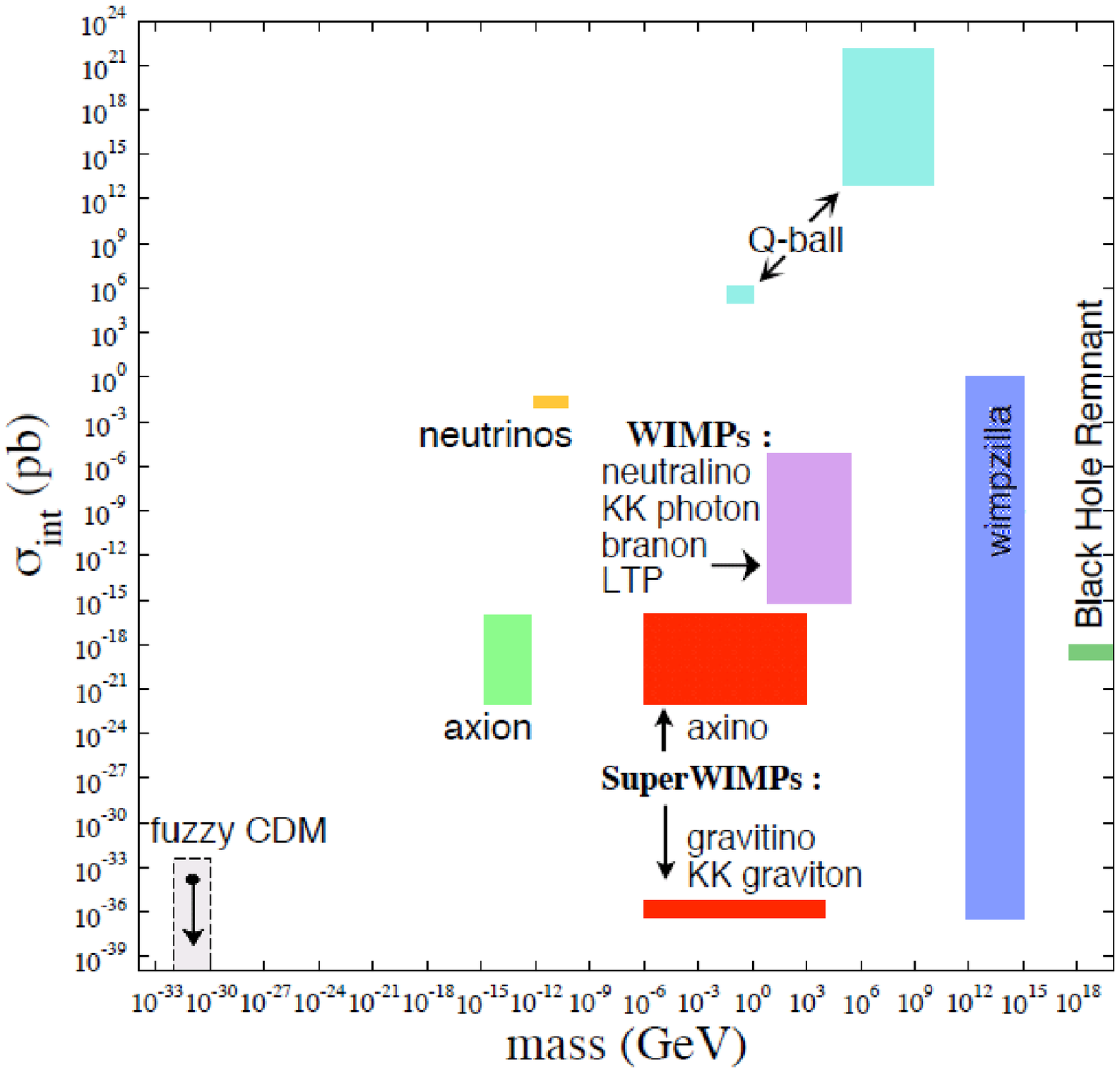}\hspace{0.1cm}
\includegraphics[width=0.54\textwidth,height=4.6cm,clip]{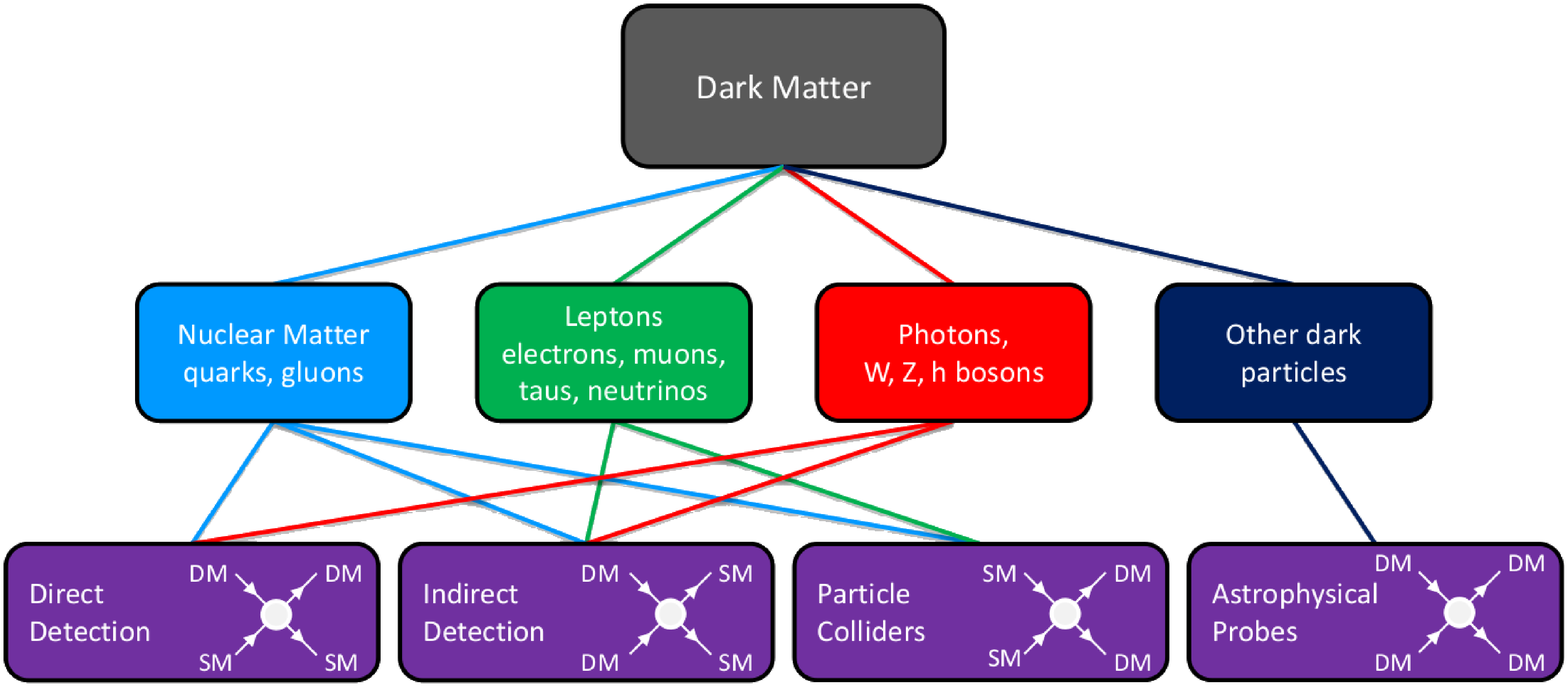}
\caption{Left: DM particle candidates in the DM-nucleon interaction cross section vs. DM-mass plane~\cite{Baer:2008uu}.
 Right: Dark matter detection approaches (depending on its interaction with SM and other DM particles): direct in underground
 experiments, indirect in space detectors, particle colliders, and astrophysical probes~\cite{Bauer:2013ihz}.}  
\label{fig:DM1}
\end{figure}

\begin{figure}
%\centering
\includegraphics[width=0.49\textwidth,height=5.4cm,clip]{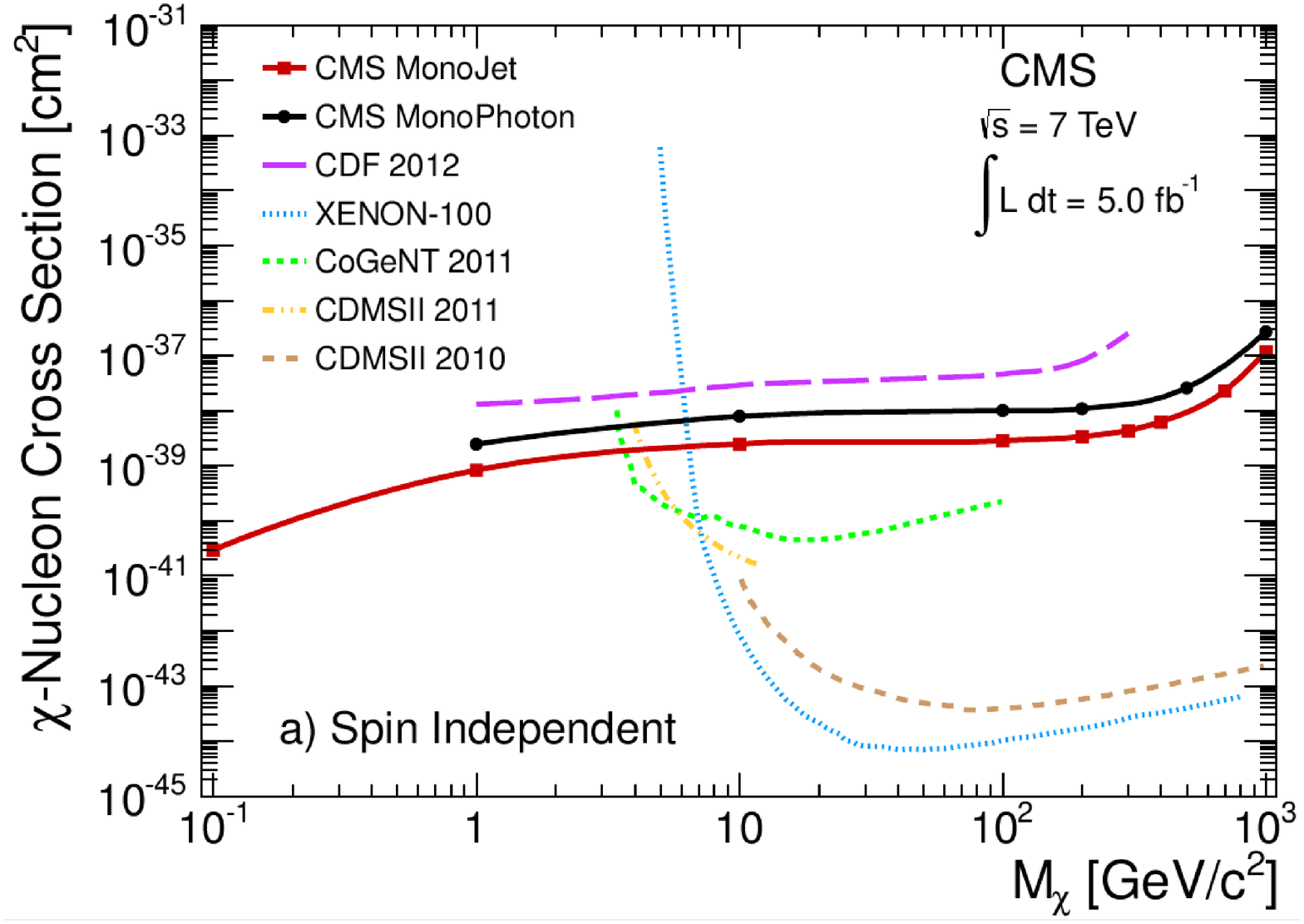}
\includegraphics[width=0.49\textwidth,clip]{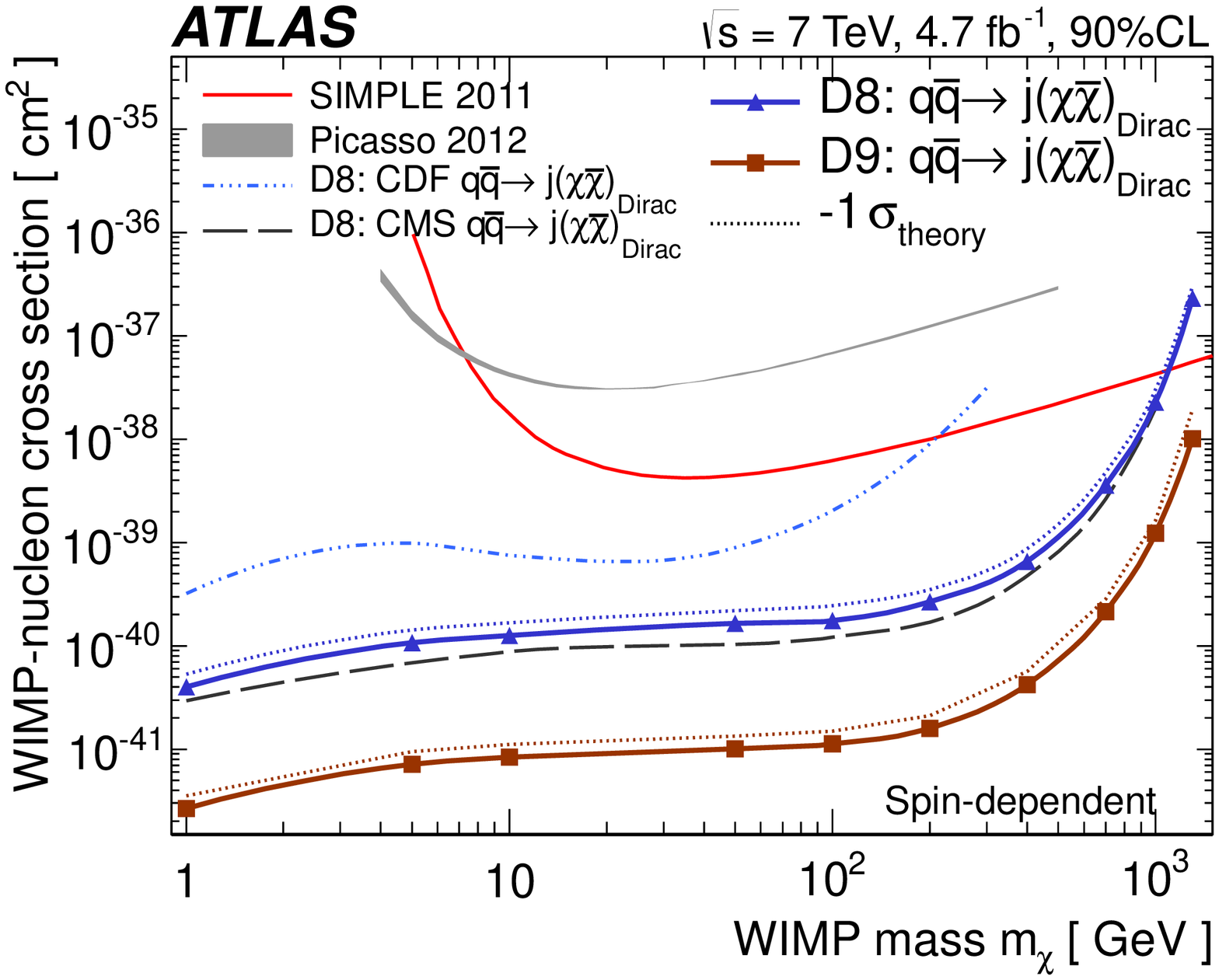} %DM_limit_sd_01.eps}
\caption{Exclusion curves in $\chi$-N interaction cross-section vs. $\chi$-mass for spin-independent (left) and
spin-dependent (right) models for various underground, space, and LHC (monojet and monophoton) searches~\cite{Chatrchyan:2012me,ATLAS:2012ky}.}
\label{fig:DM2}
\end{figure}

DM particles are gravitationally-interacting, stable, massive, cold, non baryonic, and an early universe thermal relic.
Weakly interacting massive particles (WIMP) beyond the Standard Model (SM), constitute
paradigmatic candidates with such properties. For masses m$_{\rm \chi}$~=~10 GeV--1~TeV and typical
weak interaction $\chi$-SM cross sections in the  pb-range ($\sigma_{\rm \chi-\textsc{sm}}\approx\sigma_{_{\rm EW}}$),   
one obtains the observed density $\Omega_{\rm \textsc{dm}}\propto$~m$_{\chi}^2/g_{_{\rm EW}}^4\sim$~O(10\%), 
a fact often referred to as the ``WIMP miracle''. Alternative viable possibilities exist, however, with lower
and/or higher m$_{\rm \chi}$ and $\sigma_{\rm \chi-\textsc{sm}}$  values (Fig.~\ref{fig:DM1}, left)~\cite{Baer:2008uu}. 
Popular theoretical extensions of the SM provide DM candidates such as %the lightest supersymmetric particle in 
(R-parity conserving) SUSY with a stable lightest supersymmetric particle (LSP, such as the neutralino or the
gravitino), or the lightest Kaluza-Klein tower in extra-dimensions models. Other possibilities include heavy R-handed or sterile
neutrinos, axions, or new particles of an unknown hidden sector. Apart from the astrophysical probes, three
complementary DM search strategies exist with different generic combinations of DM and SM particles in
the final- and initial-states (Fig.~\ref{fig:DM1}, right): 
\begin{itemize}
\item Direct detection in underground experiments sensitive to $\chi$-nucleon scattering ($\chi\,\rm N\to\chi\,N$):
The signals sought are anomalous nuclear recoils of order
E$_{_{\rm R}}$~=~$N\cdot\rho_{\rm \textsc{dm}}\cdot\sigma_{\rm \textsc{dm-sm}}\cdot\overline{v}_{\rm \textsc{dm}}\approx$~10~keV.
Ultralow-background scintillation/ionization/phonon detectors are required with sensitivities above 1~event/100\,kg/year.
The current limits in the WIMP-nucleon (spin-independent) interaction are approaching $\sigma_{\rm \chi-N}\approx$~10$^{-45}$~cm$^{2}$
(i.e. the zb range!) for masses m$_{\chi}\approx$~50~GeV (Fig.~\ref{fig:DM2}, left), although there have
been conflicting (non-reproducible) signals consistent with $\sigma_{\rm \chi-N}\approx$~10$^{-41}$~cm$^{2}$
at lower masses m$_{\chi}\approx$~10--50~GeV by different experiments (DAMA, CRESST) as well as not yet
statistically-significant excesses in the same region (CDMS).

\item Indirect detection in space detectors sensitive to SM-pairs from DM-DM annihilation 
($\,\chi\,\overline{\chi}\to\rm X_{\rm \textsc{sm}}\,\overline{X}_{\rm \textsc{sm}}$ and/or DM decays 
($\,\chi\to\rm X_{\rm \textsc{sm}}\,\overline{X}_{\rm \textsc{sm}}$).
%Look for SM-pairs from DM annihilation/decays:
The typical annihilation cross sections for thermal relic WIMPs are of the order of
$\sigma\cdot\overline{v}_{\chi}\approx 3\cdot 10^{-26}$~cm$^3$/s and final-states with photons
(FERMI, Veritas), neutrinos (IceCube), or cosmic rays (Pamela, AMS) are being searched.
Some excesses with respect to the SM expectations have been seen e.g. in the PAMELA and AMS cosmic-ray
positron fluxes above a few tens of GeV~\cite{Adriani:2008zr,Aguilar:2013qda}, 
%(though none in the $\overline{\rm p}$ fluxes, which c Leptophilic DM?), 
as well as in the 130-GeV line of the FERMI $\gamma$-ray fluxes from the galaxy center~\cite{Weniger:2012tx}.
%Fermi confirms excess but not DM interpretation.
DM interpretations are, however, complicated by the existing uncertainties in the astrophysical propagation
of the detected particles and/or by possible extra background sources (pulsars, unaccounted SM contributions, etc.).

\item Collider searches through DM (pair) production in the final state 
(e.g. p\,p\,$\to\chi\,\overline{\chi}+\rm X_{\rm \textsc{sm}}$).
The typical DM collider signature is large missing transverse energy (MET) from undetected LSP
%, generic WIMPs,
%or invisible Higgs decay. Almost all SUSY searches at the LHC lightest stable SUSY particle 
%(1) Lightest Particle ($\chi^0$) in RP-conserving SUSY:Prominent WIMP candidate.
coming from the decay cascade of heavier SUSY particles, (often) accompanied with multiple jets and/or leptons. 
Almost all SUSY searches at the LHC rely on extra MET from a stable $\chi$, for which no evidence has been
found so far~\cite{Chatrchyan:2013sza,Aad:2012ms}.
%No indirect chi0 signal in simplified SUSY models so far 
Similarly, the first negative results of invisible Higgs decays to DM-pairs with masses m$_{\chi}<$~m$_{_{\rm H}}$/2
--looking at excesses in the MET distribution of associated Higgs+Z-boson production-- provide limits
in the branching ratio: BR(H\,Z$\to\chi\overline{\chi}\,ll)<$~65\%~\cite{ATLAS:2013pma} and 75\%~\cite{CMS:2013yda}.
%Beyond SUSY-type signals, the
The most generic searches of DM-pair production in final states with large MET plus initial-state QCD or QED
radiation leading to unbalanced mono-jet, mono-photon (or mono-lepton) topologies, are discussed next in more detail.
%\item Astrophysical Probes: Gravitational interactions in cosmological and astrophysical domains constitute
%the only actual proof of dark matter existence to-date. The majority of searches focuse on cold and
%collisionless dark matter candidates. Any viable DM signal observed through the other three methods
%above should be ultimately consistent with the gravitational evidences.
\end{itemize}

Generic DM searches at the LHC have looked for mono-jet~\cite{Chatrchyan:2012me,ATLAS:2012ky} and
mono-photon~\cite{Chatrchyan:2012tea,Aad:2012fw} %(or mono-lepton) 
excesses above the corresponding SM backgrounds. 
The dominant SM processes producing such topologies are $Z(\nu\nu)+j,\gamma$ ($\sim$70\%) and 
$W(\nu l_{_{\rm escape}})+j,\gamma$ ($\sim$30\%), since other electroweak and QCD backgrounds can be efficiently
removed applying vetoes in isolated leptons and $\Delta\phi$ azimuthal balance cuts. The resulting
mono-$j$,$\gamma$ excesses (or lack thereof) can then be interpreted within generic effective field theories for
a contact (i.e. mediated by a heavy particle M$_{*}$) SM-DM interaction characterized by 2 parameters: 
m$_\chi$ (the mass of the DM particle, usually considered a Dirac fermion) and $\Lambda = \rm M_{*}/\sqrt{g_\chi g_{\rm q,g}}$ 
(the scale of the effective interaction, with $g_\chi$ and $g_{\rm q,g}$ the couplings of the mediator 
to the WIMP and to quark/gluons).  %(the mass of the DM particle, usually considered a Dirac fermion) 
The corresponding interaction cross sections, 
$\sigma(\chi \rm N \to \chi N)\propto \rm (g_\chi^2\,g_{\rm q,g}^2/M_{*}^4)\cdot \mu_{\rm \chi N}^2$ where
$\mu_{\rm \chi N}$ is the reduced mass of DM-nucleon system, are derived for two types of DM-SM couplings:
spin-independent (SI) and spin-dependent (SD). The current LHC exclusion limits (90\% CL) in the  
$\sigma_{\rm \chi-N}$ versus m$_\chi$ are shown in Fig.~\ref{fig:DM2} for SI (left) and SD (right) couplings
in monojet and monophoton searches~\cite{Chatrchyan:2012me,ATLAS:2012ky}. Dark matter-nucleon scattering cross
section are excluded
%above about 10$^{-42}$~(10$^{40}$)~cm$^2$ and 10$^{41}$~(10$^{39}$)~cm$^2$ for m$_\chi$
%1031042 (1:2110~40) cm2 and 2:9010~41 (2:8310~39) cm2 for a DM particle with mass 
%0.1 (100) GeV/c2 at the 90\% CL for the spin-dependent and spin-independent models, respectively.
%Best limits for low DM mass in particular for spin-dependent DM-SM couplings
above $\sigma_{\rm \chi-N}\approx$~10$^{-39}$~cm$^{2}$ (SI) and $\sigma_{\rm \chi-N}\approx$~10$^{-41}$~cm$^{2}$
(SD) for masses m$_\chi\approx$~0.1--10~GeV. Within the approximations of the underlying effective
field theory, these are the most constraining limits for a dark matter particle with mass below
m$_{\chi}$~=~3.5 GeV for spin-independent interactions, a region unexplored by direct detection 
experiments. For the spin-dependent model, these are the most stringent constraints over the
m$_{\chi}$~=~0.1--200~GeV mass range.

\section{Summary}
\label{sec:summary}

The latest developments in ultra-high-energy cosmic-rays (UHECR) and dark matter (DM) physics have been
succinctly summarized. Results from the first three years of LHC operation with proton-proton collisions up to 8~TeV
that provide new insights and constraints on the nature of both particle astrophysics phenomena, have been reviewed.\\

%the identity of the highest energy cosmic-rays on Earth and constraints on  the nature of DM, have been reviewed. 
In the UHECR domain, numerous LHC studies of the non-perturbative and semihard QCD regime have confirmed,
first, that the ``knee'' at \ECR$\sim$10$^{15.5}$~eV is not due to the production of new (unobserved) particles but to a
change from light to a more heavy CR composition~\cite{d'Enterria:2011kw}. Second, the retuning of the cosmic-rays
hadronic MCs using LHC data supports a mixed composition of proton and heavier ions ($\alpha$ particles and/or
iron) at the tail of the cosmic ray spectrum up to \ECR$\sim$10$^{20}$~eV, with reduced model uncertainties (from
$\sim$50~g/cm$^2$ to $\sim$20~g/cm$^2$ for the values of the \Xmax\ shower maximum in the
atmosphere)~\cite{Pierog:2013dya}. Upcoming LHC data studies will also help to solve the current disagreement
between the observed and predicted number of muons at ground for the highest-energy showers.\\

In the DM sector, the LHC experiments have carried out many p\,p\,$\to\chi\,\overline{\chi}+\rm X_{\rm \textsc{sm}}$ searches,
complementary to those from direct measurements in underground experiments ($\chi\,\rm N\to\chi\,N$) and indirect ones 
%, $\chi\to\rm X_{\rm \textsc{sm}}\,X_{\rm \textsc{sm}}$) 
in space detectors ($\chi\,\overline{\chi}\to\rm X_{\rm \textsc{sm}}\,\overline{X}_{\rm \textsc{sm}}$). Searches of (R-parity
conserving) SUSY at the LHC rely all in excess MET from leading DM 
candidates, such as neutralinos or gravitinos, for which no evidence has been found so far. The LHC has
provided also the first constraints on direct Higgs boson couplings to the dark sector via
H\,Z$\to\chi\overline{\chi}\,ll$. The most generic DM searches in unbalanced monojet and monophoton final-states 
--where the DM particle is produced accompanied of initial-state QCD or QED radiation-- have resulted in the  
lowest current limits for the $\chi$-N interaction cross-section,
$\sigma_{\rm \chi-N}\approx$~10$^{-39}$~cm$^{2}$ (spin independent) and 
$\sigma_{\rm \chi-N}\approx$~10$^{-41}$~cm$^{2}$ (spin-dependent), at low dark-matter masses
m$_\chi\lesssim$~10~GeV. The upcoming LHC operation in 2015, with \pp\ collisions at twice the energy of the
first run, will provide a substantial increase in the sensitivity for dark matter searches.

\paragraph{\bf Acknowledgments:}~I would like to thank the organizers of ICNFP~2013 for their kind invitation to such
interesting interdisciplinary meeting.

%%%%%%%%%%%%%%%%%%%%%%%%%%%%%%%%%%%%%%%%%%%%%%%%%%%%%%%%%%%%%%%%%%%%%%%%%%%%%%%%%%%%%%%%%%%%%%%%%%%%%%%%%
%


\begin{thebibliography}{}

\bibitem{Bauer:2013ihz}
  D.~Bauer {\it et al.}, %, J.~Buckley, M.~Cahill-Rowley, R.~Cotta, A.~Drlica-Wagner, J.~Feng, S.~Funk and J.~Hewett {\it et al.},
  %``Dark Matter in the Coming Decade: Complementary Paths to Discovery and Beyond,''
  arXiv:1305.1605 [hep-ph]
  %%CITATION = ARXIV:1305.1605;%%
\bibitem{Engel:2011zzb}
  R.~Engel, D.~Heck and T.~Pierog,
  %``Extensive air showers and hadronic interactions at high energy,''
  Ann.\ Rev.\ Nucl.\ Part.\ Sci.\  {\bf 61} (2011) 467
  %%CITATION = ARNUA,61,467;%%

\bibitem{GZK} K.~Greisen, %``End to the cosmic ray spectrum?,''
  Phys.\ Rev.\ Lett.\  {\bf 16} (1966) 748; %%CITATION = PRLTA,16,748;%%
G.~T. Zatsepin and V.~A. Kuzmin, J. Exp. Theor. Phys. Lett. 4 (1966) 78
\bibitem{Hillas:1985is} A.~M.~Hillas,
  %``The Origin of Ultrahigh-Energy Cosmic Rays,''
  Ann.\ Rev.\ Astron.\ Astrophys.\  {\bf 22} (1984) 425
  %%CITATION = ARAAA,22,425;%%

\bibitem{Werner:2005jf}
  K.~Werner, F.~-M.~Liu and T.~Pierog,
  %``Parton ladder splitting and the rapidity dependence of transverse momentum spectra in deuteron-gold collisions at RHIC,''
  Phys.\ Rev.\ C {\bf 74} (2006) 044902
%  [hep-ph/0506232].  %%CITATION = HEP-PH/0506232;%%
\bibitem{Pierog:2013ria}
  T.~Pierog, I.~.Karpenko, J.~M.~Katzy, E.~Yatsenko and K.~Werner,
  %``EPOS LHC : test of collective hadronization with LHC data,''
  arXiv:1306.0121 [hep-ph]  %%CITATION = ARXIV:1306.0121;%%
\bibitem{Kalmykov:1997te}
  N.~N.~Kalmykov, S.~S.~Ostapchenko and A.~I.~Pavlov,
  %``Quark-gluon string model and EAS simulation problems at ultra-high energies,''
  Nucl.\ Phys.\ Proc.\ Suppl.\  {\bf 52B} (1997) 17
  %%CITATION = NUPHZ,52B,17;%%
%\bibitem{Ostapchenko:2004ss}
%  S.~Ostapchenko,
%  %``QGSJET-II: Towards reliable description of very high energy hadronic interactions,''
%  Nucl.\ Phys.\ Proc.\ Suppl.\  {\bf 151} (2006) 143
%%  [hep-ph/0412332]. %%CITATION = HEP-PH/0412332;%%
\bibitem{Ostapchenko:2010vb}
  S.~Ostapchenko,
  %``Monte Carlo treatment of hadronic interactions in enhanced Pomeron scheme: I. QGSJET-II model,''
  Phys.\ Rev.\ D {\bf 83} (2011) 014018
%  [arXiv:1010.1869 [hep-ph]].  %%CITATION = ARXIV:1010.1869;%%
%\bibitem{Fletcher:1994bd}
%  R.~S.~Fletcher, T.~K.~Gaisser, P.~Lipari and T.~Stanev,
%  %``SIBYLL: An Event generator for simulation of high-energy cosmic ray cascades,''
%  Phys.\ Rev.\ D {\bf 50} (1994) 5710.  %%CITATION = PHRVA,D50,5710;%%
\bibitem{Ahn:2009wx}
  E.~-J.~Ahn, R.~Engel, T.~K.~Gaisser, P.~Lipari and T.~Stanev,
  %``Cosmic ray interaction event generator SIBYLL 2.1,''
  Phys.\ Rev.\ D {\bf 80} (2009) 094003
%  [arXiv:0906.4113 [hep-ph]]. %%CITATION = ARXIV:0906.4113;%%
\bibitem{d'Enterria:2011kw}
  D.~d'Enterria {\it et al.}, %R.~Engel, T.~Pierog, S.~Ostapchenko and K.~Werner,
  %``Constraints from the first LHC data on hadronic event generators for ultra-high energy cosmic-ray physics,''
  Astropart.\ Phys.\  {\bf 35} (2011) 98;
%  [arXiv:1101.5596 [astro-ph.HE]].
%\bibitem{d'Enterria:2011jc}
%  D.~d'Enterria, R.~Engel, T.~Pierog, S.~Ostapchenko and K.~Werner,
  %``The Strong interaction at the collider and cosmic-rays frontiers,''
  Few Body Syst.\  {\bf 53} (2012) 173
 % [arXiv:1106.2453 [hep-ph]].  %%CITATION = ARXIV:1106.2453;%%

%
\bibitem{Pierog:2013dya}
  T.~Pierog,
  %``Connecting accelerator experiments and cosmic ray showers,''
  EPJ Web Conf.\  {\bf 53} (2013) 01004
  %%CITATION = 00776,53,01004;%%
\bibitem{Giordano:2012mn}
  M.~Giordano, E.~Meggiolaro and N.~Moretti,
  %``Asymptotic Energy Dependence of Hadronic Total Cross Sections from Lattice QCD,''
JHEP {\bf 1209} (2012) 031 %[arXiv:1203.0961 [hep-ph]].
\bibitem{Antchev:2013iaa}
  G.~Antchev {\it et al.}  [TOTEM Collab.],
  %``Luminosity-independent measurements of total, elastic and inelastic cross-sections at $\sqrt{s} = 7$ TeV,''
  Europhys.\ Lett.\  {\bf 101} (2013) 21004;
  %%CITATION = EULEE,101,21004;%%
%\bibitem{Antchev:2013paa}
%  G.~Antchev {\it et al.}  [TOTEM Collab.],
  %``Luminosity-Independent Measurement of the Proton-Proton Total Cross Section at $\sqrt{s}=8$~~TeV,''
  Phys.\ Rev.\ Lett.\  {\bf 111} (2013) 012001
  %%CITATION = PRLTA,111,012001;%%
\bibitem{Aad:2011eu}
  G.~Aad {\it et al.}  [ATLAS Collab.],
  %``Measurement of the Inelastic Proton-Proton Cross-Section at $\sqrt{s}=7$ TeV with the ATLAS Detector,''
  Nature Commun.\  {\bf 2} (2011) 463
 % [arXiv:1104.0326 [hep-ex]]. %%CITATION = ARXIV:1104.0326;%%
\bibitem{Chatrchyan:2012nj}
  S.~Chatrchyan {\it et al.}  [CMS Collab.],
  %``Measurement of the inelastic proton-proton cross section at $\sqrt{s}=7$ TeV,''
  Phys.\ Lett.\ B {\bf 722} (2013) 5
%  [arXiv:1210.6718 [hep-ex]].  %%CITATION = ARXIV:1210.6718;%%
\bibitem{Abelev:2012sea}
  B.~Abelev {\it et al.}  [ALICE Collab.],
  %``Measurement of inelastic, single- and double-diffraction cross sections in proton--proton collisions at the LHC with ALICE,''
  Eur.\ Phys.\ J.\ C {\bf 73} (2013) 2456
%  [arXiv:1208.4968 [hep-ex]]. %%CITATION = ARXIV:1208.4968;%%
\bibitem{Chatrchyan:2011wm}
  S.~Chatrchyan {\it et al.}  [CMS Collab.],
  %``Measurement of energy flow at large pseudorapidities in $pp$ collisions at $\sqrt{s} = 0.9$ and 7 TeV,''
  JHEP {\bf 1111} (2011) 148
   [Erratum-ibid.\  {\bf 1202} (2012) 055];
%  [arXiv:1110.0211 [hep-ex]]. %%CITATION = ARXIV:1110.0211;%%
%\cite{Chatrchyan:2013gfi}
%\bibitem{Chatrchyan:2013gfi}
%  S.~Chatrchyan {\it et al.}  [CMS Collab.],
%``Study of the underlying event at forward rapidity in pp collisions at $\sqrt{s}$ = 0.9, 2.76, and 7TeV,'' 
JHEP {\bf 1304} (2013) 072  
%[arXiv:1302.2394 [hep-ex]].  %%CITATION = ARXIV:1302.2394;%%
\bibitem{Aaij:2012pda}
  R.~Aaij {\it et al.}  [LHCb Collab.],
  %``Measurement of the forward energy flow in $pp$ collisions at $\sqrt{s}=7$ TeV,''
  Eur.\ Phys.\ J.\ C {\bf 73} (2013) 2421
%  [arXiv:1212.4755 [hep-ex]].  %%CITATION = ARXIV:1212.4755;%%
\bibitem{Aspell:2012ux}
  G.~Antchev {\it et al.} [TOTEM Collab.],
  %``Measurement of the forward charged particle pseudorapidity density in $pp$ collisions at $\sqrt{s} = 7$ TeV with the TOTEM experiment,''
  Europhys.\ Lett.\  {\bf 98} (2012) 31002
%  [arXiv:1205.4105 [hep-ex]].  %%CITATION = ARXIV:1205.4105;%%
\bibitem{Adriani:2011nf}
  O.~Adriani {\it et al.} [LHCf Collab.],
  %``Measurement of zero degree single photon energy spectra for $\sqrt{s}$ = 7-TeV proton-proton collisions at LHC,''
  Phys.\ Lett.\ B {\bf 703} (2011) 128
%  [arXiv:1104.5294 [hep-ex]].  %%CITATION = ARXIV:1104.5294;%%
\bibitem{Auger:2012wt}
  P.~Abreu {\it et al.}  [Pierre Auger Collab.],
  %``Measurement of the proton-air cross-section at $\sqrt{s}=57$ TeV with the Pierre Auger Observatory,''
  Phys.\ Rev.\ Lett.\  {\bf 109} (2012) 062002
%  [arXiv:1208.1520 [hep-ex]]. %%CITATION = ARXIV:1208.1520;%%

%\cite{Markevitch:2003at}
\bibitem{Markevitch:2003at}
  M.Markevitch {\it et al.}, %, A.~H.~Gonzalez, D.~Clowe, A.~Vikhlinin, L.~David, W.~Forman, C.~Jones and S.~Murray {\it et al.},
  %``Direct constraints on the dark matter self-interaction cross-section from the merging galaxy cluster 1E0657-56,''
  Astrophys.\ J.\  {\bf 606} (2004) 819;
%  [astro-ph/0309303].  %%CITATION = ASTRO-PH/0309303;%%
%\cite{Clowe:2006eq}
%\bibitem{Clowe:2006eq}
  D.Clowe {\it et al.}, %, M.~Bradac, A.~H.~Gonzalez, M.~Markevitch, S.~W.~Randall, C.~Jones and D.~Zaritsky,
  %``A direct empirical proof of the existence of dark matter,''
  Astrophys.\ J.\  {\bf 648}~(2006)~109 %L109
%  [astro-ph/0608407]. %%CITATION = ASTRO-PH/0608407;%%
%\cite{vanderMarel:2012xp}
\bibitem{vanderMarel:2012xp}
  R.~P.~van der Marel {\it et al.}, %, M.~Fardal, G.~Besla, R.~L.~Beaton, S.~T.~Sohn, J.~Anderson, T.~Brown and P.~Guhathakurta,
  %``The M31 Velocity Vector. II. Radial Orbit Towards the Milky Way and Implied Local Group Mass,''
   Astrophys.\ J.\  {\bf 753} (2012) 8 %; arXiv:1205.6864 [astro-ph.GA]
  %%CITATION = ARXIV:1205.6864;%%

%\cite{Hawkins:2002sg}
\bibitem{Hawkins:2002sg}
  E.~Hawkins {\it et al.}, %S.~Maddox, S.~Cole, D.~Madgwick, P.~Norberg, J.~Peacock, I.~Baldry and C.~Baugh {\it et al.},
  %``The 2dF Galaxy Redshift Survey: Correlation functions, peculiar velocities and the matter density of the universe,''
  Mon.\ Not.\ Roy.\ Astron.\ Soc.\  {\bf 346} (2003) 78;
%  [astro-ph/0212375]. %%CITATION = ASTRO-PH/0212375;%%
%\cite{Hawkins:2002sg}
%\bibitem{Hawkins:2002sg}
  E.~Hawkins {\it et al.}, %S.~Maddox, S.~Cole, D.~Madgwick, P.~Norberg, J.~Peacock, I.~Baldry and C.~Baugh {\it et al.},
  %``The 2dF Galaxy Redshift Survey: Correlation functions, peculiar velocities and the matter density of the universe,''
  Mon.\ Not.\ Roy.\ Astron.\ Soc.\  {\bf 346} (2003) 78
%  [astro-ph/0212375].%%CITATION = ASTRO-PH/0212375;%%
%\cite{Komatsu:2010fb}
\bibitem{Komatsu:2010fb}
  E.~Komatsu {\it et al.} [WMAP Collab.],
  %``Seven-Year Wilkinson Microwave Anisotropy Probe (WMAP) Observations: Cosmological Interpretation,''
  Astrophys.\ J.\ Suppl.\  {\bf 192} (2011) 18
%  [arXiv:1001.4538 [astro-ph.CO]]. %%CITATION = ARXIV:1001.4538;%%
%\cite{Ade:2013zuv}
\bibitem{Ade:2013zuv}
  P.~A.~R.~Ade {\it et al.} [Planck Collab.],
  %``Planck 2013 results. XVI. Cosmological parameters,''
  arXiv:1303.5076 [astro-ph.CO] %%CITATION = ARXIV:1303.5076;%%
\bibitem{Baer:2008uu}
  H.~Baer and X.~Tata,
  %``Dark matter and the LHC,''
  arXiv:0805.1905 [hep-ph]. %%CITATION = ARXIV:0805.1905;%%
%\cite{Weniger:2012tx}
%\cite{Adriani:2008zr}
\bibitem{Adriani:2008zr}
  O.~Adriani {\it et al.}  [PAMELA Collab.],
  %``An anomalous positron abundance in cosmic rays with energies 1.5-100 GeV,''  
   Nature {\bf 458} (2009) 607
  %[arXiv:0810.4995 [astro-ph]].  %%CITATION = ARXIV:0810.4995;%%
%\cite{Aguilar:2013qda}
\bibitem{Aguilar:2013qda}
  M.~Aguilar {\it et al.}  [AMS Collab.], Phys.\ Rev.\ Lett.\  {\bf 110} (2013) 14,  141102
  %``First Result from the Alpha Magnetic Spectrometer on the International Space Station: Precision
  %Measurement of the Positron Fraction in Primary Cosmic Rays of 0.5~350 GeV,'' %%CITATION = PRLTA,110,141102;%%
\bibitem{Weniger:2012tx}
  C.~Weniger, JCAP {\bf 1208} (2012) 007
  %``A Tentative Gamma-Ray Line from Dark Matter Annihilation at the Fermi Large Area Telescope,''  
%  [arXiv:1204.2797 [hep-ph]].  %%CITATION = ARXIV:1204.2797;%%

%\cite{Chatrchyan:2013sza}
\bibitem{Chatrchyan:2013sza}
  S.~Chatrchyan {\it et al.}  [CMS Collab.], Phys.\ Rev.\ D {\bf 88} (2013) 052017 
  %``Interpretation of Searches for Supersymmetry with simplified Models,''  % [arXiv:1301.2175 [hep-ex]].
%\cite{Aad:2012ms}
\bibitem{Aad:2012ms}
  G.~Aad {\it et al.}  [ATLAS Collab.], Phys.\ Rev.\ D {\bf 86} (2012) 092002
  %``Further search for supersymmetry at $\sqrt{s}=7$ TeV in final states with jets, missing transverse
  %momentum and isolated leptons with the ATLAS detector,''  
  %[arXiv:1208.4688 [hep-ex]].  %%CITATION = ARXIV:1208.4688;%%

%\cite{ATLAS:2013pma}
\bibitem{ATLAS:2013pma} 
  G.~Aad {\it et al.} [ATLAS Collab.], ATLAS-CONF-2013-011
  %``Search for invisible decays of a Higgs boson produced in association with a Z boson in ATLAS,'' %%CITATION = ATLAS-CONF-2013-011;%%
%\cite{CMS:2013yda}
\bibitem{CMS:2013yda}
  S.~Chatrchyan {\it et al.}  [CMS Collab.], CMS-PAS-HIG-13-018
  %``Search for invisible Higgs produced in association with a Z boson,''
  %CITATION = CMS-PAS-HIG-13-018;%%
 
\bibitem{Chatrchyan:2012me}
  S.~Chatrchyan {\it et al.}  [CMS Collab.],
  %``Search for dark matter and large extra dimensions in monojet events in $pp$ collisions at $\sqrt{s}=7$ TeV,''
  JHEP {\bf 1209} (2012) 094
%  [arXiv:1206.5663 [hep-ex]].  %%CITATION = ARXIV:1206.5663;%%
\bibitem{ATLAS:2012ky}
  G.~Aad {\it et al.}  [ATLAS Collab.],
  %``Search for dark matter candidates and large extra dimensions in events with a jet and missing transverse momentum with the ATLAS detector,''
  JHEP {\bf 1304} (2013) 075
%  [arXiv:1210.4491 [hep-ex]]. %%CITATION = ARXIV:1210.4491;%%
\bibitem{Aad:2012fw}
  G.~Aad {\it et al.}  [ATLAS Collab.],
  %``Search for dark matter candidates and large extra dimensions in events with a photon and missing transverse momentum in $pp$ collision data at $\sqrt{s}=7$ TeV with the ATLAS detector,''
  Phys.\ Rev.\ Lett.\  {\bf 110} (2013) 011802
%  [arXiv:1209.4625 [hep-ex]]. %%CITATION = ARXIV:1209.4625;%%
\bibitem{Chatrchyan:2012tea}
  S.~Chatrchyan {\it et al.}  [CMS Collab.],
  %``Search for Dark Matter and Large Extra Dimensions in pp Collisions Yielding a Photon and Missing Transverse Energy,''
  Phys.\ Rev.\ Lett.\  {\bf 108} (2012) 261803
%  [arXiv:1204.0821 [hep-ex]]. %%CITATION = ARXIV:1204.0821;%%


%\bibitem{RefJ}
%% Format for Journal Reference
%Journal Author, Journal \textbf{Volume}, page numbers (year)
%% Format for books
%\bibitem{RefB}
%Book Author, \textit{Book title} (Publisher, place, year) page numbers
%% etc
\end{thebibliography}
\end{document}